\documentclass[epj]{svjour}
\usepackage{graphicx}
\usepackage{siunitx}
\usepackage{lineno}
\usepackage{graphicx}
\usepackage{amsmath}
\usepackage{color}
\usepackage{bm}
\usepackage{amssymb}
\usepackage{mathrsfs}
\usepackage{amsbsy}
\usepackage{caption}
\usepackage{subfigure}
\usepackage{epstopdf}

\newcommand{\be}{\begin{equation}}
\newcommand{\ee}{\end{equation}}

\numberwithin{equation}{section}
\begin{document}

\title{Analogue of the pole-skipping phenomenon in acoustic black holes}

\author{Haiming Yuan \inst{1} \and Xian-Hui Ge \inst{1,2,}\thanks{gexh@shu.edu.cn (corresponding author)}}

\institute{Department of Physics, Shanghai University, Shanghai 200444, China \and Center for Gravitation and Cosmology, College of Physical Science and Technology,Yangzhou University, Yangzhou 225009, China}

\date{Received: date / Revised version: date}

\abstract{
 The pole-skipping phenomenon is a special property of the retarded Green's function of black hole perturbations. We turn to its analog in acoustic black holes, which may relate to experiments. The frequencies of these special points are located at negative integer (imaginary) Matsubara frequencies $\omega=-i2\pi Tn$, which are consistent with the imaginary frequencies of quasinormal modes (QNMs). This implies that the lower-half plane pole-skipping phenomena have the same physical meaning as the imaginary part of QNMs, which represents the dissipation of perturbation of acoustic black holes and is related to the instability time scale of perturbation.
}
\maketitle
\section{Introduction}\label{sec:intro}

\qquad The retarded Green's function is not unique at a special point in complex momentum space $(\omega,k)$ and this phenomenon is known as ``pole-skipping'' \cite{Grozdanov1,Blake,Grozdanov2}. The retarded Green's function is given by
\be
G^R(\omega,k)_{T^{00}T^{00}}=\frac{b(\omega,k)}{a(\omega,k)}.
\ee
The location of the special points makes the coefficient $a(\omega_\star, k_\star)=b(\omega_\star, k_\star)=0$. Then, the retarded Green's function becomes $G^R(\omega_\star, k_\star)=0/0$. So if we find the intersection of zeros and poles in the retarded Green's functions, we can obtain these special points. We can use the simpler method, the AdS/CFT duality, to solve special points from the bulk field equation \cite{Makoto1,Makoto2,BlakeDavison,Makoto3}. On the bulk side, there is no unique incoming mode at the horizon, similar to the ``pole-skipping'' phenomenon in holographic chaos.\\
\indent The upper-half $\omega$-plane special point contains the information of quantum chaos. We can extract the Lyapunov exponent $\lambda$ and the butterfly velocity $v_B$ from it. Although the special points located at the lower-half $\omega$-plane are not related to the information of quantum chaos, the retarded Green's functions are also not unique at these special points. The general pole-skipping points in the lower-half $\omega$-plane are located at negative integer (imaginary) Matsubara frequencies $\mathfrak{w}_{n}=-in$ $(n=1,2\dots)$, where $\mathfrak{w}=\frac{\omega}{2\pi T}$. These special points have been found in BTZ black hole \cite{BlakeDavison},  Schwarzschild-AdS spacetime \cite{Makoto3}, 2D CFT \cite{Das}, a holographic system with the chiral anomaly \cite{Abbasi1}, a holographic system at finite chemical potential \cite{Abbasi2}, hyperbolic space \cite{Yongjun1,Kim}, the large $q$ limit of SYK chain \cite{Choi}, anisotropic plasma \cite{Karunava}, Lifshitz, and Rindler geometries \cite{Yuan}. However, the physical interpretation of the pole-skipping phenomenon remains elusive. Further investigations of the pole-skipping phenomenon in a different scenario and its possible connections to the experiments are of significant interest.\\
\indent Analogue black holes provide new windows of looking at problems between astrophysical phenomena with tabletop experiments. Using hydrodynamical flows as analogous systems to mimic a few properties of black hole physics has been proposed in \cite{Unruh}. Sound waves in a moving fluid could, in principle, analogize light waves in curved spacetime. ``Acoustic (sonic) black hole'' (ABH) shows that sound waves cannot escape from the horizon like light waves. The horizon, ergo-sphere, and Hawking radiation of $(3+1)-$dimensional static and rotating acoustic black holes have been studied in \cite{Visser1}. Some authors construct a general acoustic regular black hole that gives rise to a non-vanishing partition function that coincides with that of a conformally related black hole \cite{Lan}. Acoustic black holes for relativistic fluids can also be derived from the Abelian Higgs model \cite{Ge1,Ge2,Ge3}. The acoustic black holes might be created in high-energy physical processes \cite{Ge4}. In Ref.\cite{Ling}, the authors analyze the horizon structure of the acoustic charged black hole in curved spacetime. ABH provides a concrete laboratory model for curved space quantum field theory that we can experiment. The analogue model reflects important features of general relativity and gravity.\\
\indent Perturbations of classical gravitational backgrounds involving black holes naturally lead to quasinormal modes (QNMs) which are eigenmodes of dissipative systems. The eigenfrequencies $\omega_{QNM}$ have both a real and an imaginary part, with the real part denotes the amplitude and the imaginary part denotes the mode's damping time, which is associated with the decay timescale of the perturbation \cite{Kokkotas,Berti,Cai}.\\
\indent In \cite{Yuan}, we obtain the ``pole-skipping'' points in Rindler geometry and reveal their universality for different spacetime backgrounds. There should be universal observations; one can only use near horizon analysis to obtain ``pole-skipping'' points, not depending on the UV property of the Green's function. So we study the pole-skipping phenomenon in analogue black holes under various conditions. The frequencies of these special points are located at negative integer (imaginary) Matsubara frequencies, similar to what was obtained in \cite{Makoto1,Makoto2,BlakeDavison,Makoto3}. We combine the pole-skipping phenomenon with acoustic black holes to explore the possible realization of pole-skipping in tabletop experiments and its universality. We calculate pole-skipping points in three different acoustic black holes to compare the data under different backgrounds. We consider the backgrounds of embedding $(2+1)-$dimensional acoustic black holes into Minkowski spacetime in section~\ref{sec:Minkowski}, Schwarzschild spacetime in section~\ref{sec:general}, and AdS-Schwarzschild spacetime in section~\ref{sec:Gross}, respectively. The acoustic black hole pole-skipping in flat spacetime may be measured experimentally. However, the embedding of acoustic black holes in other curved spacetime has its own theoretical significance, which can show the universality of this phenomenon. We show that the frequencies of the lower-half plane of pole-skipping points in all cases of the acoustic black hole are consistent with the imaginary part of the frequencies of QNMs in section~\ref{sec:Discussion}, which implies that they may have the similar physical meaning as it was discussed in gravitational black holes \cite{Yuan}.
\section{$(2+1)-$dimensional acoustic black holes in Minkowski spacetime}\label{sec:Minkowski}

\begin{figure}[htbp]
    \begin{center}
   \includegraphics[width=\columnwidth]{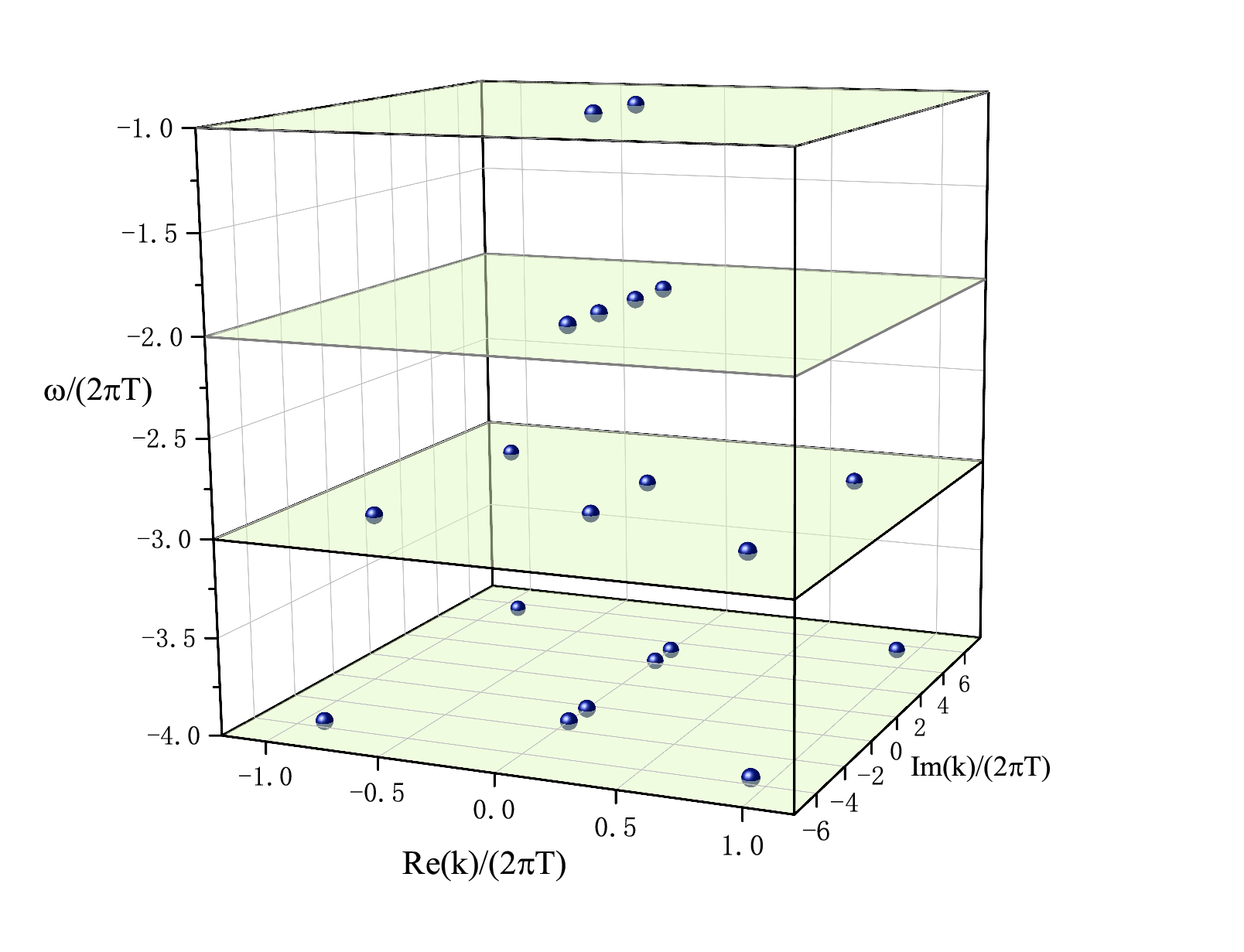}
    \caption{The pole-skipping points in acoustic black holes embedded in Minkowski spacetime.}
    \label{fig:Figure1}
    \end{center}
\end{figure}

\begin{figure}[htbp]
    \begin{center}
  \includegraphics[width=\columnwidth]{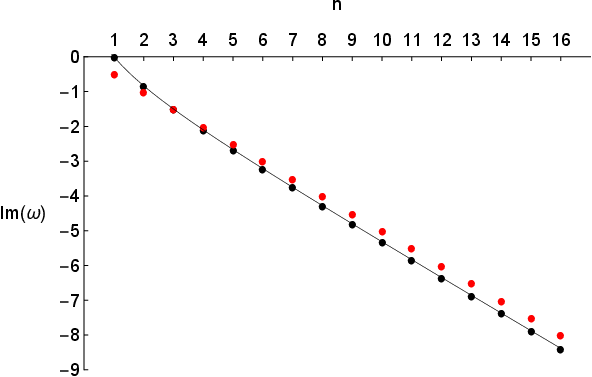}
    \caption{Red and black dots correspond to the pole-skipping points and quasinormal modes, respectively. We fit the black dots with a thin black line to make the comparison more obvious. The frequencies of pole-skipping points are close to those of quasinormal modes.\ ($c_s$=1/4)}
    \label{fig:Figure2}
    \end{center}
\end{figure}

\qquad We can derive the acoustic black hole metric from the fluid continuity equation in a uniform fluid medium \cite{Unruh,Visser1} and nonlinear Schr\"{o}dinger Equation (NSE) \cite{Marino1}. Now we show how to obtain the metric from NSE. The Nonlinear Schr\"{o}dinger Equation is given as \cite{Kivshar,Boyd}
\begin{equation}
\partial_{z}E=\frac{i}{2k}\nabla^2E-i\frac{kn_2}{n_0}E\vert E\vert^2,
\label{eq:11}
\end{equation}
where $z$ is the propagation direction, $k$ is the wave number, $n_0$ is the linear refractive index, and $E$ is the slowly varying envelope of the electromagnetic field. Substituting the complex scalar field in terms of its amplitude and phase $E=\rho^{1/2}e^{i\phi}$ into the equation \eqref{eq:11}, the hydrodynamic continuity and Euler equation become \cite{Marino1,Marino2}
\begin{align}
   \label{eq:12}
 & \partial_t\rho+\nabla\cdot(\rho\textbf{v})=0, \\
   \label{eq:13}
 &\partial_t\psi+\frac{1}{2}v^2+\frac{c^2n_2}{n^3_0}\rho-\frac{c^2}{2k^2n^2_0}\frac{\nabla^2\rho^{1/2}}{\rho^{1/2}}=0,
\end{align}
where the optical intensity $\rho$ corresponds to fluid density, and $\textbf{v}=\frac{c}{kn_0}\nabla\phi\equiv\nabla\psi$ is the fluid velocity. The dynamics take place in the transverse plane $(x,y)$ of the laser beam so that the propagation coordinate $z$ plays the role of an effective time variable $t=\frac{n_0}{c}z$. By setting $\rho=\rho_0+\epsilon\rho_1+O(\epsilon^2)$ and $\psi=\psi_0+\epsilon\psi_1+O(\epsilon^2)$, equations \eqref{eq:12} and \eqref{eq:13} can be rewritten as \cite{Marino1}
\begin{align}
     \label{eq:14}
&\partial_t\rho_1+\nabla\cdot(\rho\nabla\psi_1+\rho_1\textbf{v}_0)=0, \\
    \label{eq:15}
&\partial_t\psi_1+\nabla\psi_1\cdot\textbf{v}_0=\frac{c^2}{4k^2n^2_0}\big[\nabla\cdot\big(\frac{\nabla\rho_1}{\rho_0}\big)-\frac{\rho_1}{\rho_0}\nabla\cdot\big(\frac{\nabla\rho_0}{\rho_0}\big)\big]\nonumber\\
&-\frac{c^2n_2}{n^3_0}\rho_1,
\end{align}
When the quantum pressure is negligible, equations \eqref{eq:14} and \eqref{eq:15} can be reduced to a single second-order equation for the phase perturbations
\begin{align}
\label{eq:16}
&-\partial_t\big(\frac{\rho_0}{c^2_s}(\partial_t\psi_1+\textbf{v}_0\cdot\nabla\psi_1)\big)+\nabla\cdot\big(\rho_0\nabla\psi_1-\frac{\rho_0\textbf{v}_0}{c^2_s}(\partial_t\psi_1\nonumber\\
&+\textbf{v}_0\cdot\nabla\psi_1)\big)=0.
\end{align}
The metric become
\be
g_{\mu\nu}=\bigg(\frac{\rho_0}{c_s}\bigg)^2\left(\begin{array}{cc}
    -(c_s^2-v_0^2) & -\textbf{v}_0^\textbf{T} \\
    \textbf{v}_0 & \textbf{I} \\
\end{array}\right),
\ee
where $\textbf{I}$ is the two-dimensional identity matrix. The line element on the plane is
\be
ds^2=\bigg(\frac{\rho_0}{c_s}\bigg)^2\bigg[-(c^2_s-v_0^2)dt^2-2\textbf{v}_0dtd\textbf{x}+d\textbf{x}d\textbf{x}\bigg],
\ee
where $c^2_s$ is the local velocity of sound. We assume that the background flow is a spherically symmetric, stationary, and convergent flow, then we can define a new time \cite{Unruh}
\be
\tau=t+\int\frac{v_0dr}{c_s^2-v_0^2}.
\ee
The metric becomes
\begin{align}
\label{eq:1}
&ds^2=\bigg(\frac{\rho_0}{c_s}\bigg)^2\bigg[-(c^2_s-v_0^2)dt^2+\frac{c^2_s}{c_s^2-v_0^2}dr^2+r^2(d\theta^2\nonumber\\
&+{\rm sin}^2\theta d\phi^2)\bigg].
\end{align}
The notation of $c_s$ is the speed of sound in the fluid medium, $v_0$ is the fluid velocity, and $\rho_0$ is the fluid density. For simplicity, we assume that $c_s$ and $\rho_0$ are two constants if we assume that at some value of $r=r_h$, we have the background fluid smoothly exceeding the velocity of sound
\be
v_0=-c_s+\alpha(r-r_h)+O((r-r_h)^2).
\ee
The above metric assumes just the form it has for a Schwarzschild metric near the horizon. We choose $\theta=\frac{\pi}{2}$ for a spatially two-dimensional fluid model; then, we rewrite the $(2+1)-$dimensional acoustic black hole metric \eqref{eq:1} in the following form
\be
\label{eq:5}
ds^2=-f(r)dt^2+\frac{c_s}{f(r)}dr^2+r^2d\phi^2,
\ee
where $f(r)=2c_s\alpha(r-r_h)$, and $\alpha$ is a parameter associated with the velocity of the fluid.  $r_h$ is the location of an acoustic event horizon. $T=\frac{f'(r_h)}{4\pi\sqrt{c_s}}$ gives the Hawking temperature. In order to obtain the pole-skipping points, we use the Eddington-Finkelstein (EF) coordinates. By substituting the tortoise coordinate $dr_{\ast}=\frac{\sqrt{c_s}}{f(r)}dr$ and $v=t+r_{\ast}$ into the metric \eqref{eq:5}, we obtain
\be
ds^2=-f(r)dv^2+2\sqrt{c_s}dv dr+r^2d\phi^2,
\ee
We consider the propagation of a scalar wave of form $\psi=e^{-i\omega v+ik\phi}\psi(r)$ and submit it into the Klein-Gordon equation
\be
\partial^\mu(\sqrt{-g}g^{\mu\nu}\partial_\nu \psi)=0.
\ee
The equation becomes
\begin{align}
\label{eq:2}
&\psi''(r)+\frac{f(r)+r f'(r)-2i\omega r\sqrt{c_s}}{r f(r)}\psi'(r)\nonumber\\
&-\frac{k^2\sqrt{c_s}+i\omega r\sqrt{c_s}}{r^2f(r)}\psi(r)=0.
\end{align}
We use approximation $f(r)\sim f'(r_h)(r-r_h)$ and expand the field equation near horizon $r=r_h$
\be
\psi''(r)+(1-i\mathfrak{w})\frac{\psi'(r)}{r-r_h}+\frac{2\pi T\mathfrak{k}^2+i\mathfrak{w}r_h}{2r_h^2}\frac{\psi(r)}{r-r_h}=0,\nonumber\\
\ee
where $\mathfrak{w}=\frac{\omega}{2\pi T}$, and $\mathfrak{k}=\frac{k}{2\pi T}$. For a generic $(\mathfrak{w},\mathfrak{k})$, the equation has a regular singularity at $r=r_h$. One can solve it by a power series expansion around $r=r_h$
\be
\label{eq:3}
\psi(r)=(r-r_h)^\chi \sum^\infty_{n=0}\psi_{n}(r-r_h)^n.
\ee
At the lowest order, we can obtain the indicial equation $\chi(\chi-i\mathfrak{w})=0$. The two solutions yield
\be
\label{eq:17}
\chi_1=0,\qquad \chi_2=i\mathfrak{w}.
\ee
One corresponds to the incoming mode and another the outgoing mode. If we choose $i\mathfrak{w}=1$ and the appropriate value of $\mathfrak{k}$, make the singularity in front of $\psi'(r)$ and $\psi(r)$ terms vanishing, we call it a ``pole-skipping'' point. The regular singularity at $r=r_h$ becomes a regular point at this special point. We take the coefficients $(1-i\mathfrak{w})$ and $\frac{2\pi T\mathfrak{k}^2+i\mathfrak{w}r_h}{2r_h^2}$ to be vanishing. We then obtain the location of the special point in the $\psi(r)$ field equation
\be
\label{eq:18}
\mathfrak{w}_{\ast}=-i,\nonumber\\
\mathfrak{k}^{2}_{\ast}=-\frac{r_h}{2\pi T}.
\ee
From equation \eqref{eq:18}, two solutions in \eqref{eq:17} become
\be
\label{eq:19}
\chi_1=0,\qquad \chi_2=1.
\ee
We extend the pole-skipping phenomenon at higher Matsubara frequencies $\omega_n=-i2\pi Tn$ by using the method given in \cite{BlakeDavison}. We insert \eqref{eq:3} into \eqref{eq:2} and expand the equation of motion in powers of $(r-r_h)$. Then, a series of perturbed equations in the order of $(r-r_h)$ can be denoted as
\begin{align}
\label{eq:4}
&S=\sum^\infty_{n=0}S_n(r-r_h)^n=S_0+S_1(r-r_h)+S_2(r-r_h)^2\nonumber\\
&+\dots.
\end{align}
We write down the first few equations $S_n=0$ in the expansion of \eqref{eq:4}
\begin{align}
&0=M_{11}(\omega,k^2)\psi_{0}+(2\pi T-i\omega)\psi_{1},\nonumber\\
&0=M_{21}(\omega,k^2)\psi_{0}+M_{22}(\omega,k^2)\psi_{1}+(4\pi T-i\omega)\psi_{2},\nonumber\\
&0=M_{31}(\omega,k^2)\psi_{0}+M_{32}(\omega,k^2)\psi_{1}+M_{33}(\omega,k^2)\psi_{2}\nonumber\\
&+(6\pi T-i\omega)\psi_{3}.
\end{align}
To obtain an incoming solution, we should solve a set of linear equations of form
\begin{align}
\label{eq:7}
&\mathcal{M}^{(n)}(\omega,k^2)\cdot \psi\equiv\nonumber\\
&\left(\begin{array}{ccccc}
    M_{11} & (2\pi T-i\omega) & 0    & 0  &\dots\\
    M_{21} & M_{22}& (4\pi T-i\omega)& 0   &\dots\\
    M_{31} & M_{32}&  M_{33} &(6\pi T-i\omega) &\dots\\
    \dots   &  \dots&  \dots  &\dots   &\dots\\
\end{array}\right)\left(\begin{array}{ccccc}
    \psi_{0}\\
   \psi_{1}\\
    \psi_{2} \\
    \dots \\
\end{array}\right)\nonumber\\
&=0.
\end{align}
The locations of special points $(\omega_n, k_n)$ can be easily extracted from the determinant of the $(n\times n)$ matrix $\mathcal{M}^{(n)}(\omega,k^2)$ constructed by the first $n$ equations. The first three order pole-skipping points are shown in figure~\ref{fig:Figure1}. We have chosen $r_0=1$, $\alpha=1$, and $c_s=1/\sqrt{3}$.\\
\indent In Ref.\cite{BlakeDavison,Karunava,Yuan,Jeong}, the authors show that dispersion relation, which arises from the pole of the retarded Green's function (QNMs), passes through the lower half-plane pole-skipping points. We want to see if the relation between the pole-skipping points and the hydrodynamic dispersion relation still holds for acoustic black holes. The QNMs of Unruh's acoustic black hole are given by \cite{Saavedra}
\be
\omega_{QNM}=-\frac{i}{2}\frac{(n-1)(n+3)\alpha}{n+1}.
\ee
where $\alpha$ represents the velocity of the fluid. We compare the frequencies of pole-skipping points with those of QNMs. The ratio of frequencies is a constant $\omega_\ast/\omega_{QNM}\approx2\sqrt{c_s}$. We can normalize $\omega_{QNM}$ as $\tilde\omega_{QNM}\rightarrow 2\sqrt{c_s}\ \omega_{QNM}$. So that $\omega_\ast/\tilde\omega_{QNM}\approx1$. Under such normalization, we can see that the pole-skipping points and the QNM points are closely related to the QNMs (figure~\ref{fig:Figure2}). This, in turn, implies that lower-half plane pole-skipping points are also related to the damping of black hole QNMs.

\section{$(2+1)-$dimensional acoustic black holes embedded in Schwarzschild spacetime}\label{sec:general}

\begin{figure}[htbp]
    \begin{center}
     \includegraphics[width=\columnwidth]{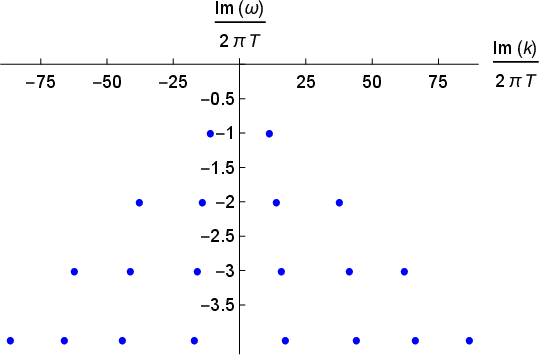}
    \caption{The pole-skipping points in acoustic black holes from general relativistic fluids near acoustic horizon $r_h$.\ ($r_h=6$)}
    \label{fig:Figure3}
    \end{center}
\end{figure}

\qquad Next, we will consider the acoustic metric in curved spacetime. This section considers that $(2+1)-$dimensional acoustic black hole metric embedded in Schwarzschild spacetime obtained from general relativistic fluids. The continuity equation is
\be
\partial_\mu\bigg(\sqrt{-g^{GR}}nu^\mu\bigg)=0,
\ee
where $n$ is the entropy density of the fluid; $u^\mu$ is the velocity of the fluid, and $u_\mu u^\mu=g^{GR}_{\mu\nu}u^\mu u^\nu=-1$. The equation describing the propagation of the linearized field $\psi$ is given in \cite{Visser2,Bil,Moncrief,Wang}
\begin{align}
\label{eq:9}
&\partial_\mu\bigg\{\frac{n}{w}\sqrt{-g^{GR}}\bigg[g^{\mu\nu}_{GR}-\bigg(1-\bigg(\frac{n}{w}\frac{\partial w}{\partial n}\bigg)^{-1}\bigg)u^\mu u^\nu\bigg]\bigg\}\partial_\nu \psi\nonumber\\
&=0,
\end{align}
where $w$ is enthalpy. By taking $c_s^2=(\frac{n}{w}\frac{\partial w}{\partial n})^{-1}\vert_{s/n}$, the form of metric can be recast as \cite{Visser2,Bil,Moncrief}
\be
\label{eq:10}
ds^2=\bigg[g^{GR}_{\mu\nu}+(1-c_s^2)u_\mu u_\nu\bigg]dx^\mu dx^\nu.
\ee
From the perspective of general relativistic fluid mechanics in the $(2+1)-$dimensional Schwarzschild background, the metric \eqref{eq:10} becomes
\begin{align}
&ds^2=\bigg(\frac{2}{3}-f_0\bigg)dt^2-\frac{9f_0 r+2r_h-6r}{6f_0 r-9f_0^2 r}dr^2\nonumber\\
&+r^2d\phi^2.
\end{align}
The radius of the acoustic horizon is $r_h=6M$, and $M$ is the mass of the black hole. We define $f(r)=\frac{2}{3}-f_0$ and $f_0=1-\frac{2M}{r}$. The Hawking temperature can be calculated as $T=\frac{1}{4\sqrt{3}r_h}$. The equation \eqref{eq:9} near the acoustic horizon $r=r_h$ by using $f(r)\sim f'(r_h)(r-r_h)$ is
\begin{align}
&\psi''(r)+(1-i\mathfrak{w})\frac{\psi'(r)}{r-r_h}-\frac{2\pi T\mathfrak{k}^2+i\mathfrak{w}r_h}{2r_h^2}\frac{\psi(r)}{r-r_h}\nonumber\\
&=0.
\end{align}
We obtain the first order special point by taking the terms $(1-i\mathfrak{w})$ and $\frac{2\pi T\mathfrak{k}^2+i\mathfrak{w}r_h}{2r_h^2}$ to vanish
\be
\mathfrak{w}_{\ast}=-i,\nonumber\\
\mathfrak{k}^{2}_{\ast}=-\frac{r_h}{2\pi T}.
\ee
We can work out the first four order pole-skipping points, as shown in figure~\ref{fig:Figure3}. Note that $\omega_\ast$ again takes the same value as the previous section. This reveals that the value of $\omega_\ast$ is universal, not only for gravitational black holes but also for acoustic black holes.

\section{$(2+1)-$dimensional acoustic black holes embedded in AdS-Schwarzschild spacetime}\label{sec:Gross}

\begin{figure}[htbp]
  \begin{subfigure}
         \centering
        \includegraphics[width=\columnwidth]{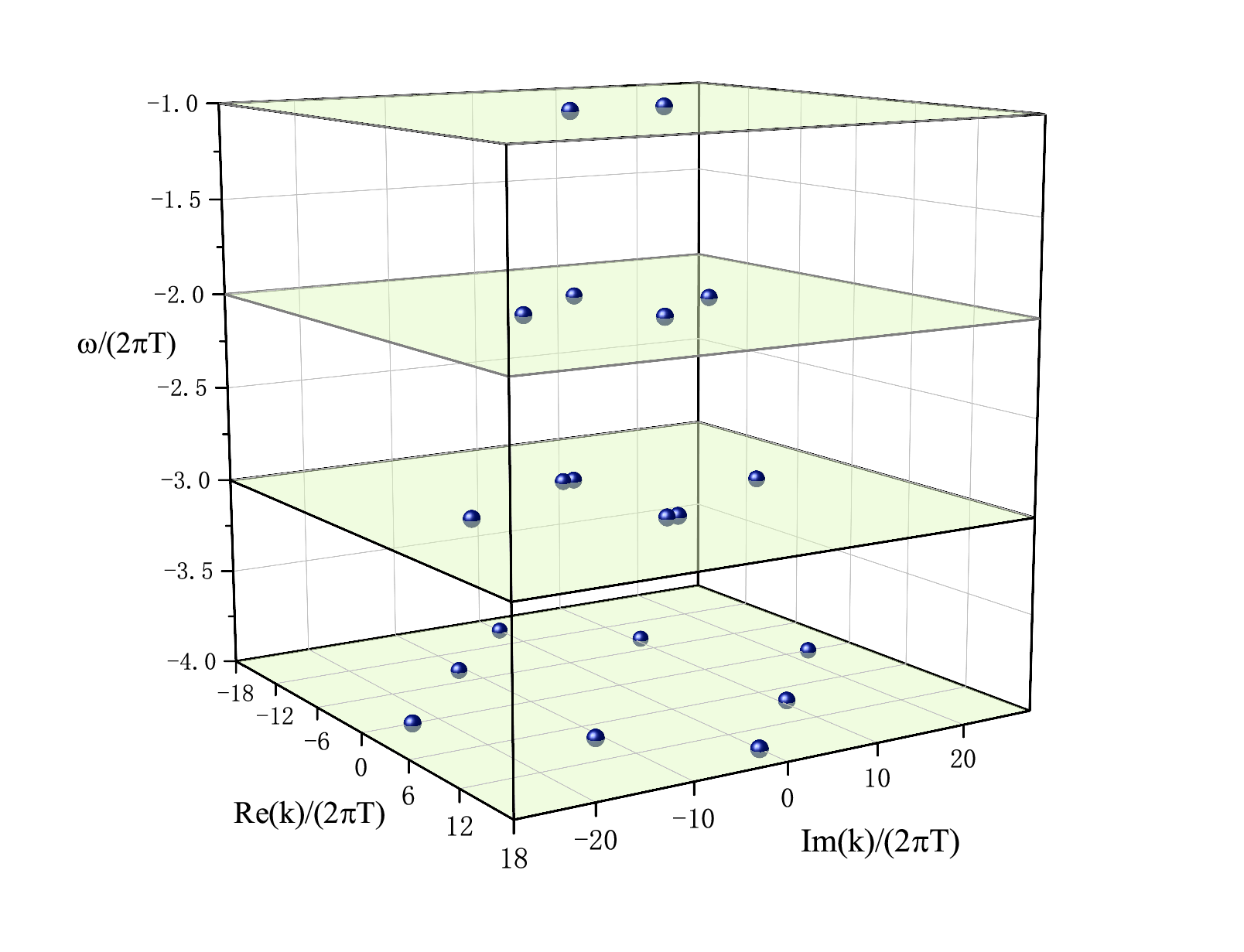}
         \caption{The pole-skipping points in acoustic black holes from Gross-Pitaeskii equation embedded in curved spacetime near acoustic horizon $r_h=\sqrt{3}r_0$.\ ($r_0=1$)}\label{fig:Figure4}
    \end{subfigure}
     \quad
     \begin{subfigure}
         \centering
        \includegraphics[width=\columnwidth]{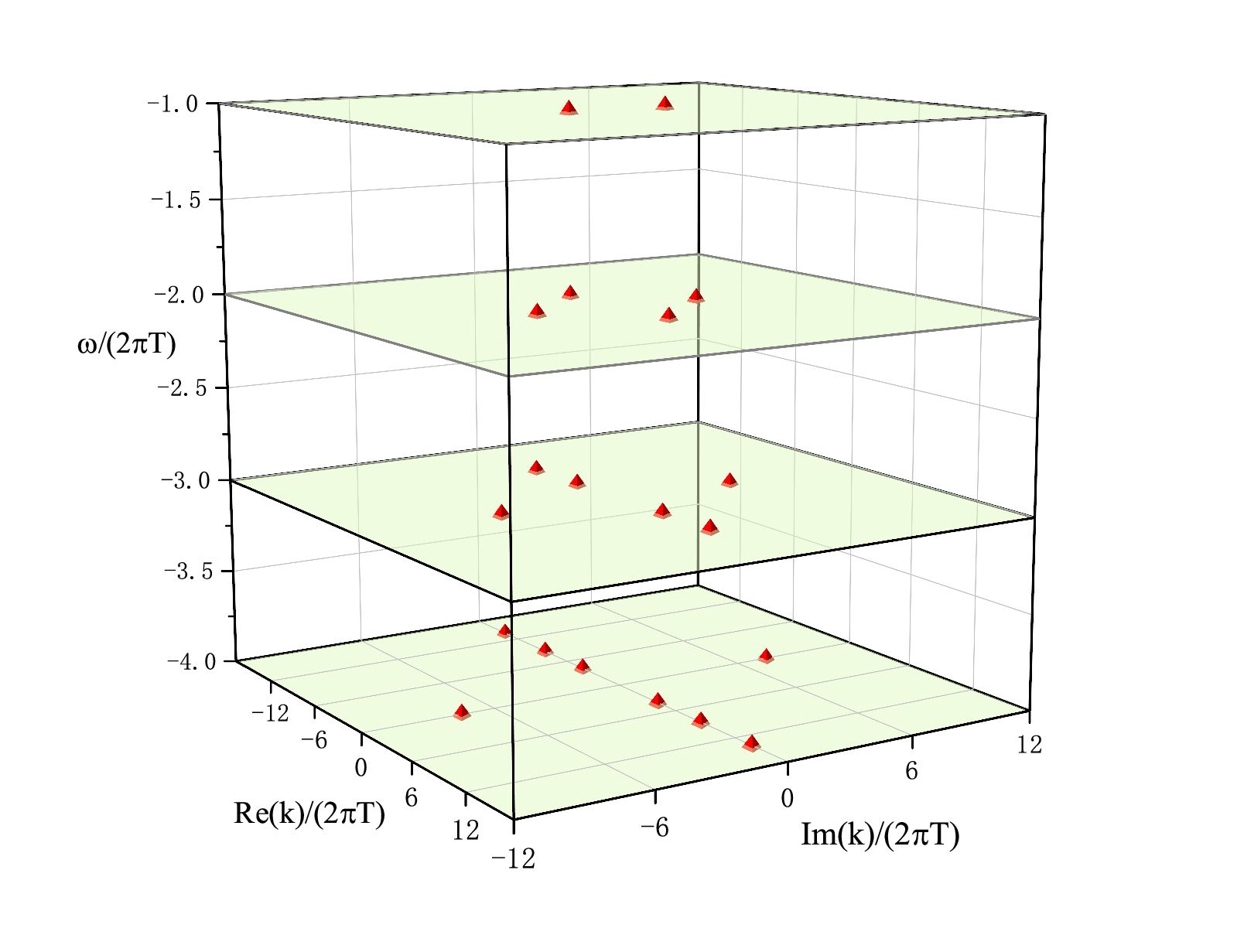}
        \caption{The pole-skipping points in acoustic black holes from Gross-Pitaeskii equation embedded in curved spacetime near event horizon $r_0$.\ ($r_0=1$)}\label{fig:Figure5}
     \end{subfigure}
 \end{figure}

\qquad We will consider a $(2+1)-$dimensional acoustic black hole metric embedded in AdS-Schwarzschild spacetime obtained from the Gross-Pitaeskii equation in this section. Compared with the cases of acoustic black holes in flat and Schwarzschild spacetime, it is interesting to consider the conditions that there are two horizons in this system. So we can find two kinds of pole-skipping points in this section. The Gross-Pitaeskii theory in curved spacetime is given by \cite{Gross,Pitaevskii}
\be
S=\int d^{4}x\sqrt{-g}(\vert\partial_\mu\varphi\vert^2+m^2\vert\varphi\vert^2-\frac{b}{2}\vert\varphi\vert^4),
\ee
where $\varphi$ is a complex scalar order parameter. Embedded $(2+1)-$dimensional acoustic black hole metric \eqref{eq:1} in AdS-Schwarzschild spacetime \cite{Ge1,Wang}, it is given by
\begin{align}
&ds^2=(g_{\mu\nu}^{GR}\star g_{\mu\nu}^{ABH})dx^\mu dx^\nu=\mathcal{G}_{tt}dt^2+\mathcal{G}_{rr}dr^2\nonumber\\
&+\mathcal{G}_{\phi\phi}d\phi^2,
\end{align}
where
\begin{align}
&\mathcal{G}_{tt}=-\frac{1}{3}f_{ABH}(r)f_{GR}(r),\nonumber\\
&\mathcal{G}_{rr}=\frac{1}{f_{ABH}(r)f_{GR}(r)},\quad \mathcal{G}_{\phi\phi}=r^4,\nonumber\\
&f_{ABH}(r)=1-\frac{r_h^2}{r^2},\quad f_{GR}(r)=r^2(1-\frac{r_0^2}{r^2}).
\end{align}
The acoustic horizon $r_h$ is located at $\sqrt{3}r_0$, which is required to be larger than the event horizon $r_0$ of the black hole. The Hawking temperature is given by
\begin{align}
&T=\frac{1}{4\pi\sqrt{\mathcal{G}_{rr}}}\big(-\sqrt{\frac{g^{ABH}_{tt}}{-g^{GR}_{tt}}}g'^{GR}_{tt}+\sqrt{\frac{-g^{GR}_{tt}}{g^{ABH}_{tt}}}g'^{ABH}_{tt}\big)\vert_{r=r_h}\nonumber\\
&=\frac{r_h^2-r_0^2}{6\pi r_0}.
\end{align}
Using the Eddington-Finkelstein (EF) coordinate, the metric can be written as
\be
ds^2=-\frac{1}{3}F(r)dv^2+\frac{2}{\sqrt{3}}dv dr+r^4d\phi^2,
\ee
where $F(r)=f_{ABH}(r)f_{GR}(r)$. The relativistic wave equation is given as \cite{Ge1}
\be
\label{eq:6}
\partial^\mu(\sqrt{-\mathcal{G}}\mathcal{G}^{\mu\nu}\partial_\nu \psi)=0.
\ee
We submit the perturbation of scalar wave $\psi=e^{-i\omega v+ik\phi}\psi(r)$ into \eqref{eq:6}
\begin{align}
\label{eq:8}
&\psi''(r)+\frac{2F(r)+r F'(r)-2\sqrt{3}i\omega r}{r F(r)}\psi'(r)\nonumber\\
&-\frac{k^2+2\sqrt{3}i\omega r^3}{r^4F(r)}\psi(r)=0.
\end{align}
\textbf{$\bullet$ Case 1: near acoustic horizon \bm{$r_h$}}\quad We use the approximation $F(r)\sim F'(r_h)(r-r_h)$ near the acoustic black hole horizon $r=r_h$
\begin{align}
&\psi''(r)+(1-i\mathfrak{w})\frac{\psi'(r)}{r-r_h}-\frac{\pi T\mathfrak{k}^2+\sqrt{3}i\mathfrak{w}r_h^3}{\sqrt{3}r_h^4}\frac{\psi(r)}{r-r_h}\nonumber\\
&=0.
\end{align}
We take the value of coefficients $(1-i\mathfrak{w})$ and $\frac{\pi T\mathfrak{k}^2+\sqrt{3}i\mathfrak{w}r_h^3}{\sqrt{3}r_h^4}$ to be 0 and eliminate the singularity in front of $\psi'(r)$ and $\psi(r)$ terms. Then, we find the location of the special point
\be
\mathfrak{w}_{\ast}=-i,\nonumber\\
\mathfrak{k}^{2}_{\ast}=-\frac{\sqrt{3}\, r_h^3}{\pi T}.
\ee
We expand the field equation near horizon $r_h=\sqrt{3}r_0$. Now we repeat the process we have shown in section~\ref{sec:Minkowski} and calculate the special points $(\omega_n, k_n)$. The first three order pole-skipping points are in figure~\ref{fig:Figure4}.\\
\textbf{$\bullet$ Case 2: near event horizon \bm{$r_0$}}\quad If we consider the pole-skipping points near event horizon $r_0$, equation \eqref{eq:9} near the event horizon $r=r_0$ by using $f(r)\sim f'(r_0)(r-r_0)$ is
\begin{align}
&\psi''(r)+(1-i\mathfrak{w})\frac{\psi'(r)}{r-r_0}-\frac{\pi T\mathfrak{k}^2+\sqrt{3}i\mathfrak{w}r_0^3}{\sqrt{3}r_0^4}\frac{\psi(r)}{r-r_0}\nonumber\\
&=0.
\end{align}
The first order special point is located at
\be
\mathfrak{w}_{\ast}=-i,\nonumber\\
\mathfrak{k}^{2}_{\ast}=-\frac{\sqrt{3}\, r_0^3}{\pi T}.
\ee
The location of the first three order pole-skipping points is shown in figure~\ref{fig:Figure5}. The $\mathfrak{k}^{2}$ of the first order pole-skipping point near acoustic horizon to that near the event horizon is $\mathfrak{k}_{\ast h}^{2}/\mathfrak{k}_{\ast 0}^{2}=r^3_h/r^3_0$. The data at these two horizons are only numerically different with which depend on $r_h$ and $r_0$. Combined with the first order pole-skipping points obtained above, we conclude that frequency $\mathfrak{w}_{\ast}$ is always located at $-i$. The square of momentum $\mathfrak{k}_{\ast}^{2}$ depends on the exponential power of the spatial metric value in the direction of momentum. For example, if the value of spatial metric $d\phi^2$ is $r^m$, then the square of momentum $\mathfrak{k}_{\ast}^{2}\sim r_{h,0}^{m-1}/T$ \,($r_{h,0}$ is acoustic/event horizon).
\section{Discussion and Conclusion}\label{sec:Discussion}

\qquad We have shown the consistency between pole-skipping points and QNMs for Unruh's acoustic black hole. Now we further conjecture that QNMs are consistent with the pole-skipping phenomenon in all cases of acoustic black holes. The analytic WKB approximations for QNMs are given as \cite{Schutz,Iyer,Cardoso}
\be
\omega_{QNM}=\Omega_c l-i(n+1/2)\vert\lambda\vert,
\ee
where $n$ is the overtone number, $\Omega_c$ is the angular velocity at the unstable null geodesic, and $l$ is the angular momentum of the perturbation. The notation $\lambda$ denotes the Lyapunov exponent. The real part of the complex QNM frequencies is found to be an integer multiple of the orbital angular frequency at the photon ring; the imaginary part is related to the Lyapunov exponent determining the unstable null orbits
at the photon ring radius \cite{Glampedakis}. If we take the value of the maximal chaotic Lyapunov exponent as $2\pi T$ near the black hole horizon, the imaginary part of QNMs becomes $-i(n+1/2)2\pi T$. The coefficient of $1/2$ plays the same role as $1/2$ of the ground state in quantum mechanics. When we drop off this constant by $1/2$, the result is consistent with the frequencies of pole-skipping points ($-in2\pi T$) for acoustic black holes. So we regard that the lower-half plane pole-skipping points in the acoustic black hole as some kind of dissipation of fluid perturbation.\\
\indent In summary, we show that pole-skipping also exists in acoustic black holes. For acoustic black holes in Minkowski space, the existence of the pole-skipping indicates that it might be verified in the experiments. For acoustic black holes embedded in Schwarzschild black hole and AdS-Schwarzschild black hole, the existence of the pole-skipping implies its universality. For all cases, the frequencies $\omega$ of these special points are the same as those black holes in gravity, which are located at negative integer (imaginary) Matsubara frequencies $\mathfrak{w}_{n}=-in$ $(n=1,2\dots)$. We conclude that the phenomenon of pole-skipping still exists in acoustic black holes, which are consistent with those in gravity.\\
\indent We cannot obtain poles-skipping points by solving Green's function directly without the boundary condition. Nevertheless, we have obtained the ``pole-skipping'' points by solving the bulk equations of motion near the horizon in Rindler geometry \cite{Yuan}. The procedure developed in \cite{Yuan} also applies to the acoustic black holes. For example, we calculate ``pole-skipping'' points in equation \eqref{eq:18} near the acoustic horizon in section~\ref{sec:Minkowski}. Two incoming waves \eqref{eq:17} at these ``pole-skipping'' points correspond to the nonuniqueness of the Green's function on the boundary. Combining with the previous conclusion in Rindler geometry, we conclude that the pole-skipping phenomenon does not depend on the UV property of the Green's function.\\
\indent We have shown that the frequencies of lower-half plane pole-skipping points in an acoustic black hole are the same as the imaginary frequencies of QNMs. This conclusion is consistent with our results in gravitational black holes \cite{Yuan}. This implies that the lower-half plane pole-skipping phenomenon has the same physical meaning as the imaginary of QNMs, which represents the dissipation of perturbation of acoustic black holes and is related to the decay timescale of the perturbation.

\section*{Acknowledgements}
We would like to thank Matteo Baggioli, Yu-Qi Lei, and Qing-Bing Wang for helpful discussions. This work is partly supported by NSFC (No.11875184).\\

\appendix

\section{Details of near-horizon expansions}
In this appendix, we show the details of the near-horizon expansions of the equations of motion.
\subsection{Acoustic black holes in Minkowski spacetime}
\qquad We can calculate a Taylor series solution to the equation of field $\psi(r)$ \eqref{eq:2} when the matrix equation \eqref{eq:7} is satisfied. The first few elements of this matrix are shown below
\begin{align}
&M_{11}=-\frac{1}{2r_h^2}[k^2+i\omega r_h],\nonumber\\
&M_{21}=-\frac{i\omega}{4r_h^2},\nonumber\\
&M_{22}=\frac{1}{4r_h^2}[-k^2+12\pi T r_h-5i\omega r_h+r_h^2f''(r_h)],\nonumber\\
&M_{31}=0,\nonumber\\
&M_{32}=\frac{1}{12r_h^2}[16\pi T-6i\omega+5r_h f''(r_h)+r_0^2f^{(3)}(r_h)],\nonumber\\
&M_{33}=\frac{1}{6r_h^2}[-k^2+40\pi T r_h-9i\omega r_h+3r_h^2f''(r_h)],\nonumber\\
&M_{41}=0,\nonumber\\
&M_{42}=\frac{1}{48r_h^2}[9f''(r_h)+7r_hf^{(3)}(r_h)+r_h^2f^{(4)}(r_h)],\nonumber\\
&M_{43}=\frac{1}{24r_h^2}[72\pi T-15i\omega+21r_hf''(r_h)+4r_h^2f^{(3)}(r_h)],\nonumber\\
&M_{44}=\frac{1}{8r_h^2}[84r_h\pi T-13r_hi\omega+6r_h^2f''(r_h)-k^2].
\end{align}
\subsection{Acoustic black holes embedded in Schwarzschild spacetime}
\qquad We can calculate a Taylor series solution to the equation of field $\psi(r)$ \eqref{eq:9} when the matrix equation \eqref{eq:7} is satisfied. The first few elements of this matrix are
\begin{align}
&M_{11}=-\frac{1}{2r_h^2}[k^2+r_h i\omega],\nonumber\\
&M_{21}=-\frac{i\omega}{4r_h^2},\nonumber\\
&M_{22}=\frac{1}{4r_h^2}[-k^2+12\pi Tr_h-5ir_H\omega+r_h^2f''(r_h)],\nonumber\\
&M_{31}=0,\nonumber\\
&M_{32}=\frac{1}{12r_h^2}[16\pi T-6i\omega+5r_hf''(r_h)+r_h^2f^{(3)}(r_h)],\nonumber\\
&M_{33}=-\frac{1}{6r_h^2}[k^2+9r_hi\omega-40r_h\pi T-3r_h^2f''(r_h)],\nonumber\\
&M_{41}=0,\nonumber\\
&M_{42}=\frac{1}{48r_h^2}[9f''(r_h)+7r_hf^{(3)}(r_h)+r_h^2f^{(4)}(r_h)],\nonumber\\
&M_{43}=\frac{1}{24r_h^2}[72\pi T-15i\omega+21r_hf''(r_h)+4r_h^2f^{(3)}(r_h)],\nonumber\\
&M_{44}=\frac{1}{8r_h^2}[84r_h\pi T-13r_hi\omega+6r_h^2f''(r_h)-k^2].
\end{align}
\subsection{Acoustic black holes embedded in AdS-Schwarzschild spacetime}
\qquad We can calculate a Taylor series solution to the equation of field $\psi(r)$ \eqref{eq:8} when the matrix equation \eqref{eq:7} is satisfied. The first few elements of this matrix are ($r_H$ is acoustic/event horizon)
\begin{align}
&M_{11}=-\frac{1}{6r_H^4}[\sqrt{3}k^2+6i\omega r_H^3],\nonumber\\
&M_{21}=-\frac{3i\omega}{2r_H^2},\nonumber\\
&M_{22}=\frac{1}{12r_H^4}[-\sqrt{3}k^2+6r_H^3(12\pi T-5i\omega)+\sqrt{3}r_H^4F''(r_H)],\nonumber\\
&M_{31}=-\frac{i\omega}{r_H^3},\nonumber\\
&M_{32}=\frac{1}{36r_H^4}[288\pi T r_H^2-108i\omega r_H^2+10\sqrt{3}r_H^3F''(r_H)\nonumber\\
&+\sqrt{3}r_H^4F^{(3)}(r_H)],\nonumber\\
&M_{33}=\frac{1}{18r_H^4}[-\sqrt{3}k^2+240\pi T r_H^3-54i\omega r_H^3\nonumber\\
&+3\sqrt{3}r_H^4F''(r_H)],\nonumber\\
&M_{41}=-3i\omega/2,\nonumber\\
&M_{42}=\frac{1}{144r_H^4}[720\pi T-252i\omega+54\sqrt{3}r_HF''(r_H)\nonumber\\
&+14\sqrt{3}r_H^2F^{(3)}(r_H)+\sqrt{3}r_H^3f^{(4)}(r_H)],\nonumber\\
&M_{43}=\frac{1}{72r_H^4}[2r_H^2\{648\pi T-135i\omega+21\sqrt{3}r_HF''(r_H)\nonumber\\
&+2\sqrt{3}r_H^2F^{(3)}(r_H)\}],\nonumber\\
&M_{44}=\frac{1}{24r_H^4}[6r_H^3\{84\pi T-13i\omega+\sqrt{3}r_HF''(r_H)\}\nonumber\\
&-\sqrt{3}k^2].
\end{align}

\end{document}